\documentclass[a4paper,12pt]{article}

\usepackage{graphicx,amsmath,color,cite}

\newcommand{\openone}{1\!\!1}

\newcommand{\tr}{\mathrm{tr}}

\newcommand{\E}{\mathcal{E}}
\newcommand{\mc}{\mathcal}
\newcommand{\dd}{{\mathrm d}}

\begin{document}

\title{\bf Gaussian Matrix Product States}
\author{Norbert Schuch, Michael M.\ Wolf, and J.\ Ignacio Cirac}
\date{Max-Planck-Institut f\"ur Quantenoptik,\\
Hans-Kopfermann-Str.\ 1, D-85748 Garching, Germany.}
\maketitle

\begin{center}
{\large\bf Abstract}\\[0.5cm]
\parbox{0.9\textwidth}{
We introduce Gaussian Matrix Product States (GMPS), a generalization of
Matrix Product States (MPS) to lattices of harmonic oscillators. Our
definition resembles the interpretation of MPS in terms of projected
maximally entangled pairs, starting from which we derive several
properties of GMPS, often in close analogy to the finite dimensional case:
We show how to approximate arbitrary Gaussian states by MPS, we discuss
how the entanglement in the bonds can be bounded, we demonstrate how the
correlation functions can be computed from the GMPS representation, and
that they decay exponentially in one dimension, and finally relate GMPS
and ground states of local Hamiltonians.\\[0.6cm]
\textbf{\footnotesize[This work originally appeared as Sec.~VII of
quant-ph/0509166, and is published in the Proceedings on the conference on
Quantum information and many body quantum systems, edited by M.~Ericsson
and S.~Montangero, pg.~129 (Edizioni della Normale, Pisa, 2008).]}
}
\end{center}

\hspace{1.2cm}

\section{Introduction}

In the last years, there has been considerable activity on the border
between quantum information theory and condensed matter physics, and
quantum information concepts have successfully been applied to the
description of quantum many-body systems. An important step has been the
interpretation of Matrix Product States (MPS) in terms of projected
maximally entangled pairs. MPS form a hierarchy of states which prove very
successful as a variational ansatz for simulating ground states of 
one-dimensional quantum systems, as done in the Density Matrix
Renormalization Group (DMRG) method~\cite{white:DMRG-PRL,schollwoeck:rmp}.
From the perspective of quantum information, MPS are formed by taking
virtual maximally entangled pairs between adjacent sites and applying a
linear map on each site to obtain the physical
system~\cite{frank:dmrg-mps}. This entanglement-based description 
led to a better understanding of MPS and gave rise to new
algorithms and extensions of DMRG to e.g.\ thermal states, time
evolutions, and higher dimensional
systems~\cite{guifre:timeevol,VGC04,frank:2D-dmrg,vidal:mera},
but also to new analytical tools for investigating e.g.\ quantum phase
transitions, renormalization group transformations, or the sequential
generation of quantum
states~\cite{wolf:mps-qpt,frank:renorm-MPS,schoen:hen-and-egg}.

Given the success of the MPS framework in the description of
finite-dimensional spin systems, it is natural to look for generalizations
to e.g.\ bosonic or fermionic systems.  In this paper, we
introduce bosonic Gaussian Matrix Product States
(GMPS), which describe Gaussian states on lattices of
harmonic oscillators (i.e., bosonic modes). Such systems are frequently
realized in physical setups, e.g.\ by the vibrational modes of ions in
linear traps or by arrays of nanomechanical oscillators, and since they
are typically goverened by quadratic Hamiltonians, their ground and
thermal states are Gaussian.
 
Our definition of GMPS resembles the quantum information perspective on
MPS, where one takes maximally entangled pairs and applies a linear map to
obtain the physical system.  Starting from this definition, we
show that every (translation
invariant) Gaussian state can be represented as a (translation invariant)
GMPS,  and discuss how to minimize the amount of entanglement used in the
bonds -- different from the finite-dimensional case, this is an issue
since bosonic bonds can carry an unbonded amount of entanglement.  We
discuss the properties of two-point correlation functions of GMPS and show
that they can be easily computed from the GMPS representation;  for the
case of pure one-dimensional GMPS, we prove that the correlations decay
exponentially (as it is the case in finite dimensions) and explicitly
derive the correlation length.  We end our discussion on GMPS by showing
that -- again in analogy to the finite dimensional case -- every GMPS is
the ground state of a local Hamiltonian. 

Since Gaussian states are completely characerized by their second moments
and thus by a number of parameters quadratic in the system size, unlike
for spin systems Gaussian MPS will not by themselves yield an
exponentially more efficient parametrization.  However, they can be used
to describe translational invariant states with a constant number
of parameters and thus also in the limit of an infinite chain.
Moreover, the GMPS parametrization should have favorable properties 
e.g.\ for variational minimizations with respect to local observables.

The results presented in this work have already been applied
in~\cite{AE06,AE07}.

\section{Gaussian states\label{sec:gaussian-states}}

Consider a system of $N$ bosonic modes which are characterized by $N$
pairs of canonical operators $(Q_1,P_1,\ldots, Q_N ,P_N)=:R$, and where
the canonical commutation relations (CCR) are governed by the symplectic
matrix $\sigma$ via
\[ 
\big[R_k,R_l\big]=i\sigma_{kl}\;,\quad
\sigma=\bigoplus_{n=1}^N\left(\begin{array}{cc}0&1\\-1&0\end{array}\right)\;.
\]
Then, \emph{Gaussian states} are defined as states which have a Gaussian
Wigner distribution in phase space. Those state are frequently met in
physics, since
ground or thermal states of quadratic Hamiltonians are Gaussian states,
and evolution under a quadratic Hamiltonian leaves Gaussian states
Gaussian.
They are completely characterized by their first moments
$d_k=\tr{\big[\rho R_k\big]}$ (which can be changed by local operations, 
so that we set them to zero w.l.o.g.)
 and their \emph{covariance matrix} (CM)
\begin{equation}
\gamma_{kl}=\tr\Big[\rho\big\{R_k-d_k,R_l-d_l\big\}_+\Big]\;,
\label{eq:basics:def-CM}
\end{equation}
where $\{\cdot,\cdot\}_+$ is the anticommutator. The CM satisfies
$\gamma\ge i\sigma$, which expresses Heisenberg's uncertainty
relation and is equivalent to the positivity of the corresponding
density operator $\rho\geq 0$. Purity of the state is characterized by
$\det\gamma=1$ or equivalently $(\sigma\gamma)^2=-\openone$.

When the state under consideration is translational invariant, it often
proves convenient to describe it in the Fourier basis. For
simplicity, let us consider a one-dimensional translational invariant
chain of length $N$
with periodic boundaries and reorder the canonical operators such that
$R=(Q_1,\ldots,Q_N,P_1,\ldots,P_N)$.
Then,
\[
\gamma=\left(\begin{array}{cc}\gamma_Q&\gamma_{QP}\\
    \gamma_{QP}^T&\gamma_P \end{array}\right)\ ,
\]
and translational invariance is reflected by the fact that any matrix
element $A_{kl}$ (where $A\in\{\gamma_Q,\gamma_P,\gamma_{QP}\}$) depends only on
the distance $k-l$ (which is understood $\mathrm{mod}\ N$), 
so we can write $A_{k-l}:=A_{kl}$. 
  Matrices of this type are called \emph{circulant}
and are simultaneously diagonalized by the Fourier transform. 
We write the diagonal elements of the Fourier transform as
a function of the angle $\phi=2\pi m/N$ for $m=0,\dots,N-1$; the Fourier
transform of a cirulant matrix $A$ then reads $\hat A(\phi)=\sum
A_{n}e^{-in\phi}$.

An interesting property of translational invariant pure states with one
mode per site is that they are point symmetric~\cite{cmp}. This can
readily be seen from the representation 
$$
\gamma=\left(\begin{array}{cc}
    \gamma_Q&\gamma_{QP}\\\gamma_{QP}^T&\gamma_P\end{array}\right)
=\left(\begin{array}{cc}X&XY\\YX\ &\ X^{-1}+YXY\end{array}\right)
$$
in $Q$-$P$ partitioning with $X>0$, $Y=Y^T$ real~\cite{WGKWC03}, 
since $X$ and $Y$ have to be circulant and therefore commute. Hence,
$\gamma_{QP}=XY=YX=\gamma_{QP}^T$, i.e., $\gamma$ is point symmetric. Note
that this implies in particular that the Fourier transform
$\hat\gamma(\phi)$ is real.

\section{Gaussian Matrix Product States
    \label{sec:MPS:definition}}

\begin{figure}[b]
\hspace*{\fill}
\includegraphics[height=3cm]{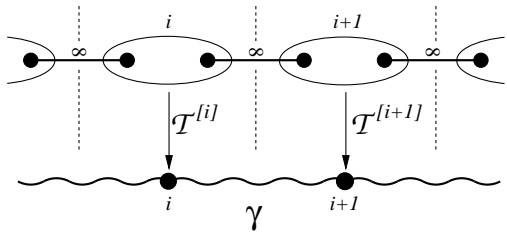}
\hspace*{\fill}
\caption{
\label{fig:GMPS:operational-def}
Construction of Gaussian Matrix Product States (GMPS). GMPS are
obtained by taking a fixed number $M$ of maximally entangled (i.e., EPR) 
pairs shared by adjacent sites, and applying an arbitrary 
$2M$ to $1$ mode Gaussian operation $\mc T^{[i]}$ on site $i$.}
\end{figure}

In the following, we introduce Gaussian matrix product states (GMPS). The definition
resembles the physical interpretation of finite-dimensional matrix product
states as projected entangled pairs: In finite dimensions, MPS can
be described by taking maximally entangled pairs of dimension $D$ between
adjacent sites and applying arbitrary local operations on each site,
thus mapping the $D\times D$ dimensional input (the virtual system) 
to a $d$-dimensional output state (the physical system).
Similarly, GMPS are obtained by taking a number of entangled bonds and
applying local (not necessarily trace-preserving) operations $\mc
T^{[i]}$, where  the boundary conditions can be taken either open or
closed.  Any GMPS is completely described by the type of
the bonds and by the operations $\mc T^{[i]}$.  Note that this
construction holds independent of the spatial dimension. For one
dimension, it is illustrated in Fig.~\ref{fig:GMPS:operational-def}.
As matrix product states are frequently used to describe translationally
invariant systems, an inportant case is given if all maps are identical,
$\mc T^{[i]}=\mc T\ \forall i$.

In order to define MPS in the Gaussian world, we have to
decide on the type of the bonds as well as on the type of operations.
We choose both the bonds to be Gaussian states and the operations to
be Gaussian operations, i.e., operations mapping Gaussian inputs to
Gaussian outputs.  For now, we will take the bonds to be maximally
entangled (i.e., EPR) states, such that the only parameter 
originating from the bonds is the number $M$ of EPRs.  
We show later on how the case of finitely entangled bonds can be easily
embedded.

\begin{figure}
\begin{center}
\includegraphics[width=4cm]{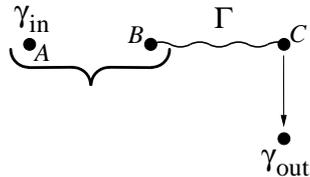}
\end{center}
\caption{
\label{fig:GMPS:jamiolkowski}
The Jamiolkowski isomorphism. The Gaussian channel described by
the state $\Gamma$ can be implemented by projecting the input state
$\gamma_\mathrm{in}$ (mode $A$) and the input port of $\Gamma$ (mode $B$) 
onto the EPR state (symbolized by curly brackets). In case of success, the
output is obtained in mode $C$.  The operation can be made trace-preserving
by measuring in a basis of displaced EPR states, and displacing $C$
according to the measurement outcome.
}
\end{figure}

As to the operations, we will allow for arbitrary Gaussian operations.
Operations of this type are most easily described by the Jamiolkowski
isomorphism~\cite{Jam72}.  There, any Gaussian operation $\mc T$ which maps
$N$ input modes to $M$ output modes can be described by an $N+M$ mode
covariance matrix $\Gamma$ with block $B$ (input) and $C$ (output).  The
corresponding map on some input state $\gamma_{\mathrm{in}}$ in mode $A$
is implemented by projecting the modes $A$ and $B$ onto an EPR state as
shown in Fig.~\ref{fig:GMPS:jamiolkowski}, such that the output state $\mc
T(\gamma_{\mathrm{in}})$ is obtained in mode $C$.  Conversely, the
matrix $\Gamma$ which represents the channel $\mc T$ is obtained by
applying the channel to one half of a maximally entangled state. The
duality between $\mc T$ and $\Gamma$  is most easily understood in terms
of teleportation, and shows that this characterization encompasses all
Gaussian operations. Note that the protocol of
Fig.~\ref{fig:GMPS:jamiolkowski} can be always made trace-preserving by
projecting onto the set of phase-space displaced EPR states and correcting the
displacement of mode $C$ according to the measurement outcome~\cite{GC02}.

In the following, we will denote all maps $\mc T$ by their corresponding
CM $\Gamma$.  Sometimes, we will speak of the modes $B$ and $C$ as input
and output ports of $\Gamma$, respectively.

We now discuss how the covariance matrix of the output will
depend on the CM of the input and on the channel
$\Gamma$~\cite{GC02,Fiu02b}.  This is most easily computed in the
framework of characteristic functions~\cite{Hol82}.
The characteristic function of the output is given by
$$
\chi_C(\xi_C)\propto\int e^{-\xi_A^T\gamma_{\mathrm{in}}\xi_A}
    e^{-\xi_{BC}^T\Gamma\xi_{BC}}
    \delta(x_A-x_B)\delta(p_A+p_B)\mathrm{d}\xi_{AB}\ ,
$$
and by integrating over subsystem $A$, we obtain
$$
\chi_C(\xi_C)\propto\int e^{-\xi_{BC}^T M \xi_{BC}}\mathrm{d}\xi_B
$$
with
$$
M=\left(\begin{array}{cc}
    \theta\gamma\theta+\Gamma_B&\Gamma_{BC}\\\Gamma_{CB}&\Gamma_C
    \end{array}\right)\ .
$$
Basically, the integration
$\int\dd\xi_A\delta(x_A-x_B)\delta(p_A+p_B)$ does the following:
first, it applies the partial transposition 
$\theta\equiv\left(\begin{smallmatrix}1&0\\0&-1\end{smallmatrix}\right)$
to one of the subsystems, and second, it collapses the two systems $A$
and $B$ in the covariance matrix by adding the corresponding entries.  The
integration over $\xi_B$, on the other hand, leads to a state whose CM is
the Schur complement of $M_{11}$, $M_{22}-M_{21}M_{11}^{-1}M_{12}$, such
that the output state is described by the CM
$$
\gamma_{\mathrm{out}}=\Gamma_C-
    \Gamma_{CB}\frac{1}{\Gamma_B+\theta\gamma_{\mathrm{in}}\theta}\Gamma_{BC}\ .
$$

Let us briefly summarize how to perform projective measurements onto the
EPR state in the framework of CMs, where we denote the measured modes by $A$
and $B$, while $C$ is the remaining part of the system. First, apply the
partial transposition to $B$, second, collapse $A$ and $B$, and third,
take the Schur complement of the collapsed mode $AB$, which gives the
output CM of $C$.

In analogy to the finite-dimensional case, we will focus on pure GMPS.
Particularly, a GMPS is pure if the $\Gamma^{[i]}$ which describe the
operations $\mc T^{[i]}$ are taken to be pure, which we assume from now
on.  Let us finally emphasize that the given defintion of MPS holds
independent of the spatial dimension of the system, as do most of the
following results, and in fact applies to an arbitrary graph.

\section{Completeness of Gaussian MPS
\label{sec:MPS:completeness}}

In the following, we show that any pure and translational invariant
state can be approximated arbitrarily well by translational invariant
Gaussian matrix product states, i.e., GMPS with identical local operations
$\mc T$.  (Without translational invariance, this is clear anyway:
the complete state is prepared locally and teleported to its
destination using the bonds.) The proof is presented for one dimension,
but can be extended to higher spatial dimensions. Note that a similar
result also holds for finite-dimensional MPS~\cite{mps-reps}.

Given a translational invariant state $\gamma$, there is a
translational invariant Hamiltonian $H$ which transforms the separable
state $\openone$ into $\gamma$, $\gamma=SS^T$, $S=e^{\sigma H}$. 
It has been shown~\cite{kraus:tinv-sim}
that this time
evolution can be approximated arbitrarily well by a sequence of
translational invariant local (one-mode) and nearest neighbor (two-mode)
Hamiltonians $H_j$,
\begin{equation}
\label{eq:GMPS:trotter-decomp}
e^{\sigma H}\approx\prod_{j=1}^J e^{\bigoplus_n\sigma H_j}\ ,
\end{equation}
where the $H_j$ act on one or two modes, respectively, and approach the
identity for growing $J$.

\begin{figure}[t]
\hspace*{\fill}
\parbox{6cm}{\includegraphics[width=6cm]{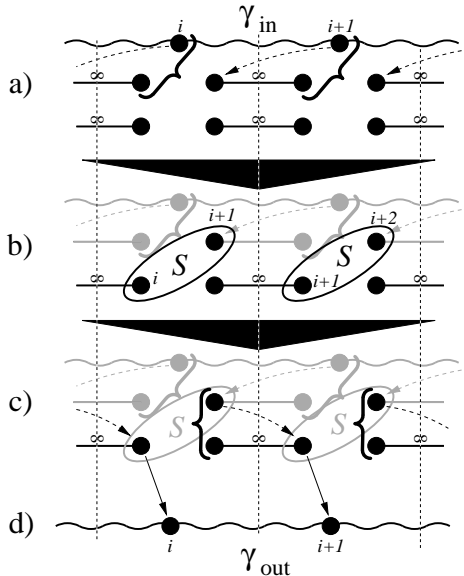}}
\hspace*{\fill}
\parbox{6cm}{
\caption{
\label{fig:GMPS:completeness}
Implementation of a translational invariant nearest neighbor Hamiltonian
in a translational invariant fashion. Starting from $\gamma_\mathrm{in}$, 
the input is first teleported to the left, then, the infinitesimal time
evolution $S=e^{\sigma H}$, $H\ll1$, is performed, and finally, the state
is teleported back.
}}
\hspace*{\fill}
\end{figure}

Clearly, translational invariant local Hamiltionians can be implemented by
local maps without using any EPR bonds. In the following, we show how
translational invariant nearest-neighbor interactions can be implemented
by exploiting the entanglement of the bonds. The whole procedure is
illustrated in Fig.~\ref{fig:GMPS:completeness} and requires two EPR
pairs per site. We start with some initial state $\gamma_\mathrm{in}$ onto
which we want to apply $S_\oplus=e^{\bigoplus\sigma H_j}
    \approx\openone+\bigoplus_n\sigma H_j$.

First, we perform local EPR measurements between the modes of
$\gamma_\mathrm{in}$ and one of the bonds in order to teleport the modes
of $\gamma_\mathrm{in}$ to the left,
cf.~Fig.~\ref{fig:GMPS:completeness}a.  Then, the infinitesimal symplectic
operation $S=e^{\sigma H_j}$ is applied to the left-teleported mode and
the
second bond, Fig.~\ref{fig:GMPS:completeness}b.  In the last step,
another EPR measurement is performed which teleports the left-teleported
mode back to the right, and ``into'' the mode on which the adjacent $S$
was applied.  As the operations $e^{\sigma H_j}\approx\openone+\sigma H_j$
all commute, the ``nested'' application of the nearest neighbor symplectic
operations $S$ indeed give $S_\oplus$,  and thus the remaining mode really 
contains the output
$\gamma_\mathrm{out}=S_\oplus\gamma_\mathrm{in}S_\oplus^T$.  The whole
decomposition (\ref{eq:GMPS:trotter-decomp}) can be implemented by
iterated application of the whole protocol of Fig.~\ref{fig:GMPS:completeness}.  

\section{GMPS with finitely entangled bonds}

Let us now consider the entanglement contained in the bonds and show that
infinitely entangled bonds can be replaced by finitely entangled ones.
Intuitively, this should be possible whenever the channel $\mc T^{[i]}$
destroys some of the entanglement of the bond anyway, i.e., $\Gamma^{[i]}$
is non-maximally entangled. In that case, it should be possible to use a
less entangled bond while choosing a channel which does not destroy
entanglement any more.

\begin{figure}[t]
\hspace*{\fill}
\includegraphics[width=13cm]{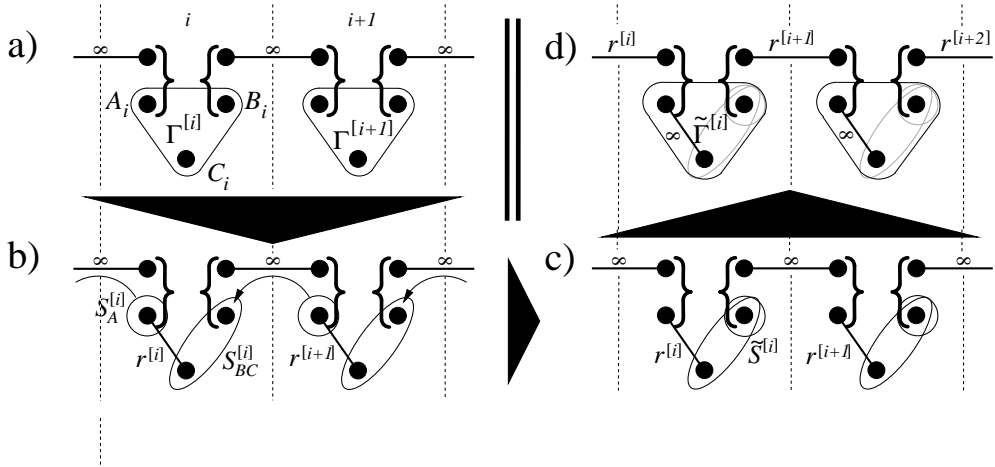}
\hspace*{\fill}
\caption{
\label{fig:GMPS:nonmax-ent}
How to make the bonds of GMPS finitely entangled. \textbf{a)} The initial
MPS.  \textbf{b)} Do a Schmidt decomposition of the original map $\Gamma$.
\textbf{c)} Move the $S^{[i]}_A$ through the infinitely entangled bond to
the next site.  \textbf{d)} Swap the finitely and the infinitely entangled
pair.
}
\end{figure}

The method is illustrated in Fig.~\ref{fig:GMPS:nonmax-ent}. Again, for
reasons of clarity we restrict to one dimension and one bond. The
argument however applies independent of the spatial dimension and the
number of bonds. The only restriction we have to make is the restriction
to pure GMPS, i.e., those with pure $\Gamma^{[i]}$.

Consider a GMPS with local channels given by $\Gamma^{[i]}$ and infinitely
entangled bonds, Fig.~\ref{fig:GMPS:nonmax-ent}a. First, apply a Schmidt
decomposition~\cite{HW01} to $\Gamma^{[i]}$ in the partition $A|BC$, which
can be always done as long as $\Gamma^{[i]}$ is pure. The Schmidt
decomposition allows us to rewrite the state as shown in
Fig.~\ref{fig:GMPS:nonmax-ent}b -- an entangled state between modes $A$ and
$C$ with two-mode squeezing $r^{[i]}$, $B$ in the coherent state
$\openone$, and sympectic operations $S_A^{[i]}$ and $S_{BC}^{[i]}$ which
are applied to modes $A$ and $BC$, respectively. As the bond itself is
infinitely entangled, we can teleport the sympectic operation through the
bond to the next site as
indicated in Fig.~\ref{fig:GMPS:nonmax-ent}b. Then, $S_A^{[i+1]}$ can be
merged with $S_{BC}^{[i]}$ to a new operation $\tilde S^{[i]}$ acting on
modes $B$ and $C$ of site $i$ (Fig.~\ref{fig:GMPS:nonmax-ent}c). Finally, 
in the triples consisting of one maximally entangled state, one non-maximally
entangled state, and the projection onto the EPR state, the maximally and
the non-maximally entangled state can be swapped, resulting in
Fig.~\ref{fig:GMPS:nonmax-ent}d. There, we have finitely entangled bonds,
while the infinite entanglement has been moved into the new maps
$\tilde\Gamma^{[i]}$.

It is tempting to apply this construction to the completeness proof of the
preceding section in order to obtain a construction which is less wasting
with respect to resources. However, for any iterative protocol this is
most likely difficult to achieve. The reason for this is found in the
no-distillation theorem which states that with Gaussian operations, it is
not possible to increase the amount of entanglement between two
parties~\cite{GC02}.  Particularly, this implies that in each step of an
iterative protocol, the bonds need to have at least as much entanglement
as can be obtained at the output of this step, maximized over all inputs
where the entanglement is increased. This is indeed a severe restriction,
although it does not imply the impossibility of such a protocol. One
could, e.g., create a highly entangled state in the first step and then
approach the desired state by decreasing the entanglement in each step.
Still, it seems most likely that a sequence of MPS which approach a given
state efficiently will have to involve more and more bonds simultaneously
and thus cannot be constructed in an iterative manner.

\section{Correlation functions of Gaussian MPS}

In this section, we show how to compute correlation functions from the
maps $\Gamma^{[i]}$ which describe the GMPS. We show that this can be done
efficiently, i.e., in a time which is polynomial in the systems size,
independent of the dimension of the graph. This is different from the
finite dimensional case, where correlation functions of e.g.\
two-dimensional MPS cannot be computed efficiently~\cite{cplx-of-PEPS}.  Of
course, this is not too surprising given that Gaussian states can be fully
characterized by a number of paramaters quadratic in the number of modes.

\begin{figure}[t]
\begin{center}
\includegraphics[width=7cm]{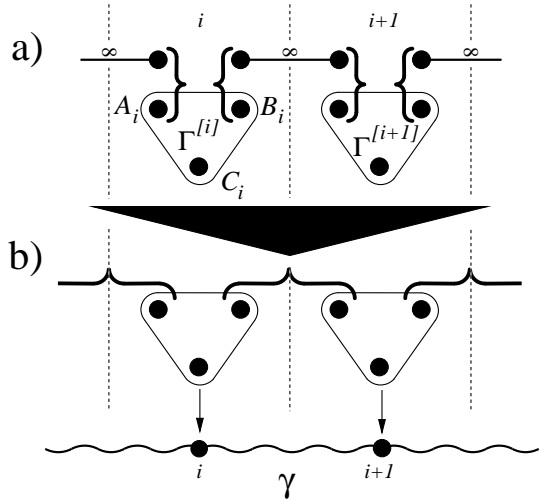}
\end{center}
\caption{If the local operations are described by states $\Gamma^{[i]}$ via
the Jamiolkowski isomorphism, the construction of GMPS can be simplified
by replacing the measurement-bond-measurement triples by a simple
projection onto the EPR state.
\label{fig:GMPS:mps-definition-with-jamiolkowski}
}
\end{figure}

Let us start with the general case of different $\Gamma^{[i]}$, as in
Fig.~\ref{fig:GMPS:mps-definition-with-jamiolkowski}a. The calculation can
be facilitated by the simple observation that the triples consisting of
two projective measurements and one EPR pair can be replaced by a single
projection onto the EPR state,
Fig.~\ref{fig:GMPS:mps-definition-with-jamiolkowski}b. It follows that we 
can apply the formalism for projective measurements onto the EPR
state which we presented in Section~\ref{sec:MPS:definition}. 
Starting from $\bigoplus_i \Gamma^{[i]}$,
we first partially transpose all $B$ modes, then collapse $A_{i+1}$ and
$B_i$ for all $i$, and finally take the Schur complement of the merged
mode. In case of periodic boundary conditions, this can be expressed by
the transformation matrix 
\begin{equation}
\label{eq:GMPS:M-Matrix}
\Pi=\left(\begin{array}{ccc}
    \openone_A & \mathcal R\theta_B & 0 \\
    0 & 0 & \openone_C 
  \end{array}\right)
\end{equation}
which maps $ABC$ onto $A'C$, where $\theta_B\equiv\theta\otimes\openone$
is the partial transposition on system $B$, and $\mathcal R$ is the
circulant right shift operator, 
$(\mathcal R)_{ij}=\delta_{i,j+1\mod N}\otimes\openone$.  Then, the
output state, i.e., the GMPS characterized by $\Gamma^{[i]}$, is
$$
\gamma=\mathrm{SC}_{A'}\left[\Pi\left(
	    \bigoplus_i\Gamma^{[i]}\right)\Pi^T\right]\ ,
$$
where $\mathrm{SC}_{X}(U)$ is the Schur complement of the $X$ part of $U$, 
$\mathrm{SC}_{X}(U)=U_{YY}-U_{YX}U_{XX}^{-1}U_{XY}$. For open boundary
conditions, the matrix $\Pi$ has to be modified accordingly at the
boundaries. All the involved operations scale polynomially in the product
$NM$ of the number of sites $N$ and the number of modes $M$.

In case all the local maps are chosen equal,
$\Gamma^{[i]}\equiv\Gamma\ \forall i$, the above formula can be simplified
considerably. 
 Therefore, note that the Fourier transform can be taken into the
Schur complement, and that $\Pi$ as well as $\bigoplus_{i=1}^N
\Gamma^{[i]}=\Gamma\otimes\openone_N$ are blockwise circulant so that both are
diagonalized by the Fourier transform. 
Thereby,
$\Gamma\otimes\openone$ is mapped onto the constant function
$\Gamma$, and the same holds for
$\openone$ and $\theta$ in (\ref{eq:GMPS:M-Matrix}). The right shift
operator $\mc R$, on the other hand, is transformed to $e^{i\phi}$\openone: 
the EPR measurement performed between adjacent sites leads to a complex
phase of $\phi$. Altogether, we have
$$
\hat{\Pi}=\left(\begin{array}{ccc}
    \openone_A & e^{i\phi}\theta_B & 0 \\
    0 & 0 & \openone_C 
  \end{array}\right)\ ;\quad
\hat\gamma=\mathrm{SC}_{A'}\left[\hat{\Pi}\,\Gamma\,\hat{\Pi}^\dagger\right]\ .
$$
Directly expressed in terms of the map $\Gamma$, this reads
\begin{equation}
\hat\gamma(\phi)=\Gamma_C-\Gamma_{C|AB}\,\hat\Lambda
    \frac{1}{\hat\Lambda\,\Gamma_{AB|AB}\,\hat\Lambda^\dagger}\,
	\hat\Lambda^\dagger\,\Gamma_{AB|C}
\label{eq:MPS:characterization}
\end{equation}
where $\hat\Lambda=(\openone_A\,;\,e^{i\phi}\theta_B)$ is 
the upper left subblock of $\hat{\Pi}$.

\section{States with rational trigonometric functions as Fourier
transforms}

Let us now restrict to pure MPS (i.e.,  those for which $\Gamma$ is
pure) with one mode per site.  As we have shown in
Section~\ref{sec:gaussian-states}, those states have reflection symmetry
and therefore
$\hat\gamma(\phi)=\gamma_0+2\sum_{n\ge0}\gamma_n\cos(n\phi)$ is real.
This implies that the sines in (\ref{eq:MPS:characterization}) can only
appear in even powers $\sin^{2n}\phi=(1-\cos^2\phi)^n$. Therefore, the
Fourier transform $\hat\gamma$ of any pure Gaussian MPS, which is a
$2\times 2$ matrix valued function of $\phi$, has elements which are
rational functons of $\cos(\phi)$,
$(\hat\gamma(\phi))_{xy}=p_{xy}(\cos(\phi))/q_{xy}(\cos(\phi))$ with $p$,
$q$ polynomials. The degree of the polynomials is limited by the 
size of $\hat\Lambda\Gamma_{AB}\hat\Lambda^\dagger$, and thus by the
number $M$ of the bonds. One can easily check that $\dim p\le2M+1$ and
$\dim q\le2M$.

For the following discussion, let us write those rational functions with a
common denominator $d$,
\begin{equation}
\label{sachertorte}
\hat\gamma(\phi)=\frac{1}{d(\cos(\phi))}
    \left(\begin{array}{cc}
	q(\cos(\phi)) & r(\cos(\phi)) \\ r(\cos(\phi)) & p(\cos(\phi)) 
    \end{array}\right)\ ,
\end{equation}
where $q$, $p$, $r$, and $d$ are polynomials of degree $L$. Then, 
the set of all such $\hat\gamma$ with $L\ge2M+1$ encompasses the set of
translational invariant GMPS with $M$ bonds. Computing correlation
functions in a lattice of size $N$ can be done straightforwardly in this
representation by taking the discrete Fourier transform of
$\hat\gamma(\phi)$ which scales polynomially with $N$, and in the
following section we show that for one dimension, the correlations can be
even computed exactly in the limit of an infinite chain.

It is interesting to note that $\gamma(\phi)$ is already determined 
up to a finite number of possibilities by fixing $r$ and $d$. Since
$\gamma$ is pure, $1=\det\gamma=\det\hat\gamma$, 
and therefore,
$pq=d^2+r^2$. Therefore, the zeros of $pq$ are the zeros of $d^2+r^2$, 
such that the only freedom is to choose how to distribute the zeros on $p$
and $q$. On the contrary, fixing only $q$ and $d$ does not give sufficient
information, while choosing $p$, $q$ and $d$ (i.e., the diagonal of
$\hat\gamma$) does not ensure that there
exists a polynomial $r$ such that $pq-r^2=d^2$.

From the above, it follows that $2L+1$ parameters are sufficient to
describe $\hat\gamma(\phi)$, where $L$ is still the degree of the
polynomials. This encloses all translational invariant Gaussian MPS with
bond number $M\le(L-1)/2$, which need $(2M+1)(2M+2)=L(L+1)$ parameters.
Therefore, the class of states where $\hat\gamma(\phi)$ is a rational
function of $\cos(\phi)$ is a more efficient description of
translationally invariant states than Gaussian MPS are.

Let us stress once more that the results of this section hold 
for arbitrary spatial dimension.

\section{Correlation length}

In the following, we show that the correlations of one-dimensional GMPS
decay exponentially, and we explicitly derive the correlation length.  The
derivation only makes use of the representation (\ref{sachertorte}) of
Gaussian MPS and thus holds for the whole class of states where the
Fourier transform is a rational function of the cosine. We will restrict
to the case where the state $\Gamma$ associated to the GMPS map  has 
no diverging 
entries, which corresponds to the case where the denominator
$d(\cos(\phi))$ in (\ref{sachertorte}) has no zero on the unit circle.%
\footnote{
The case where $d$ has zeros on the unit circle corresponds to critical
systems, which is why the correlations diverge. In the case of a
Hamiltonian $H=V\oplus\openone$, however, the ground state correlations of
$P$ do not diverge~\cite{cmp}. As in that case one has $pq=d^2$, 
$p/d=d/q$ need not have a singularity just because $q/d$ has one.}

The
correlations are directly obtained by back-transforming the elements of
$\hat\gamma(\phi)$, which are rational functions
$[\hat\gamma(\phi)]_s=s(\cos(\phi))/d(\cos(\phi))$, $s\in\{p,q,r\}$; 
in the limit of an infinite chain,
$$
(\gamma_s)_n=\frac{1}{2\pi}\int_0^{2\pi}
    \frac{s(\cos(\phi))}{d(\cos(\phi))}e^{in\phi}\dd\phi\ .
$$
Now transform $s$, $d$ to complex polynomials via
$\cos(\phi)\rightarrow(z+1/z)/2$, and  expand with $z^K$, $\tilde
s(z):=z^Ks(z)$, $\tilde d(z):=z^Kd(z)$, where $K$ is chosen large
enough to make $\tilde s$, $\tilde d$ polynomials in $z$. Then,
\begin{eqnarray*}
(\gamma_s)_n&=&\frac{1}{2\pi i}\int_{\mc S^1}
    \frac{\tilde s(z)z^{n-1}}{\tilde d(z)}\dd z\\
&=&\sum_{z_i:\tilde d(z_i)=0}\frac{1}{(\nu_i-1)!}
    \underbrace{
    \left.\frac{\dd^{\nu_i-1}}{\dd z^{\nu_i-1}}
    \left[\frac{\tilde s(z)z^{n-1}}{\tilde d_i(z)}\right]
    \right|_{z=z_i}}_{D_i}
\end{eqnarray*}
by the calculus of residues, where $\nu_i$ is the order of the zero $z_i$
in $\tilde d$ and $\tilde d_i(z)(z-z_i)^{\nu_i}=\tilde d(z)$.  For
$n>\nu_i$, $D_i\propto z_i^{(n-\nu_i)}$, and it follows that the
correlations decay exponentially, where the correlation length is given
by the largest zero of $q(z)$ inside the unit circle.

This proof only holds for one-dimensional GMPS. However, it can be proven
for arbitrary spatial dimensions that the correlations decay 
faster than any polynomial by iterated integration by parts with respect
to one component of $\phi$, 
cf.~\cite{cmp}.

\section{Gaussian MPS as ground states of local Hamiltonians}

Finally, let us focus on the relation of translational invariant 
Gaussian MPS and local
Hamiltonians.  We prove that every GMPS is the ground state of a local
Hamiltonian, while conversely most Hamiltonians do not have GMPS as an
exact ground state -- again, this is in close analogy to the
finite-dimensional case~\cite{mps-reps}.  Once more, the proof 
only requires the state to be of the form Eq.~(\ref{sachertorte}). We will
make use of some results on ground states of translational invariant
quadratic Hamiltonians presented in~\cite{cmp}.  Define the 
Hamiltonian matrix $H\ge0$ via the Hamilton operator $\mathcal H$ 
by virtue of $\mathcal H=\sum
H_{kl}R_kR_l$, as well as the spectral function $\mathcal
E=\sqrt{\mathrm{det}\,\hat H}$.  Then, the ground state is given by
\begin{equation}
\label{gamma-eq}
\hat\gamma=(\mathcal E\oplus\mathcal E)^{-1}\sigma\hat H\sigma^T
\end{equation}
and has
energy $\tfrac12\tr\,\mathcal E$.

Given a pure state $\gamma$ with Fourier transform (\ref{sachertorte}),
define 
\begin{equation}
\label{eq:GMPS:local-hamil-gs}
\hat H(\phi)=
    \left(\begin{array}{cc}
	p(\cos(\phi)) & -r(\cos(\phi)) \\ -r(\cos(\phi)) & q(\cos(\phi)) 
    \end{array}\right)
\end{equation}
Then, $H$ corresponds to a local Hamiltonian -- the interaction range is
the degree of $p,q,r$ -- and
$\E(\phi)=\big[\sqrt{pq-r^2}\;\big](\cos\phi)=d(\cos\phi)$, which together
with (\ref{gamma-eq}) proves that $\gamma$ is the ground state of $H$.

Let us also have a brief look at the converse question: Given a local
Hamiltonian, when will it have a GMPS as its ground state?
Any local and translational invariant 
Hamiltonian has a Fourier transform which consists of
polynomials in $\cos(\phi)$, and thus we adapt the notation of
Eq.~(\ref{eq:GMPS:local-hamil-gs}). Then, 
following Eq.~(\ref{gamma-eq})
the ground state is represented
by a rational function of $\cos(\phi)$ in Fourier space exactly if 
$pq-r^2=d^2$ is the square of another polynomial.
In terms of the
original Hamiltonian, this implies that
$H_QH_P-H_{QP}^2$ has to be the square of another banded matrix. 
For example, for $H=V\oplus\openone$ one would need
$V=X^2$ with $X$ again a banded matrix~\cite{CE05}.

\section*{Acknowledgements}

This work has been supported by the EU project COVAQIAL.

\end{document}